\documentclass[aps,prb,twocolumn,showpacs,longbibliography]{revtex4-1}
\usepackage{amssymb}
\usepackage{amsmath}
\usepackage{amsfonts}
\usepackage{graphicx}
\usepackage{color}
\usepackage[utf8x]{inputenx}
\usepackage{hyperref}
\usepackage{bm}

\begin{document}
\title{Optical transitions in two-dimensional topological insulators with point defects}

\author{Vladimir~A.~Sablikov}
\author{Aleksei~A. Sukhanov}
\affiliation{V.A. Kotel’nikov Institute of Radio Engineering and Electronics, Russian Academy of Sciences,
Fryazino, Moscow District, 141190, Russia}

\begin{abstract}
Nontrivial properties of electronic states in topological insulators are inherent not only to the surface and boundary states, but to bound states localized at structure defects as well. We clarify how the unusual properties of the defect-induced bound states are manifested in optical absorption spectra in two-dimensional topological insulators. The calculations are carried out for defects with short-range potential. We find that the defects give rise to the appearance of specific features in the absorption spectrum, which are an inherent property of topological insulators. They have the form of two or three absorption peaks that are due to intracenter transitions between electron-like and hole-like bound states.
\end{abstract}


\maketitle

\section{Introduction}
\label{Intro}
Topological quantum states arising in topological insulators (TIs) due to strong spin-orbit interaction and time reversal symmetry attract a great deal of interest because of their nontrivial properties as well as because they provide us with opportunities to explore qualitatively new physical phenomena challenging for construction of novel quantum devices, spintronic applications and topological quantum computation~\cite{HasanKaneRMP2010,QiZhangRMP2010,Ortmann2015topological}. The topological states are robust since they are protected against the scattering by weak non-magnetic impurities and disorders. The main attention is paid to topological states that exist near the surface of three-dimensional TIs and the edge of two-dimensional (2D) TIs. However, unusual electron states arise also at impurities and structure defects located in the bulk.

Impurity induced states were studied both for impurities on the surface of three-dimensional TIs~\cite{Zhou,Guo,Wang,Biswas,Black-Schaffer1,Black-Schaffer2}, and in 2D TIs~\cite{Shan,J_Lu,shen2013topological,Lee,SablikovPSSR2014}. In addition, the impurity states were discussed for one-dimensional topological systems~\cite{shen2013topological}. The main conclusion is that the electron density localizes near the defect in a highly unusual way. A similar phenomenon can occur in a trivial one-dimensional system with spin-orbit interaction in the presence of a weak magnetic field~\cite{GambettaPRB2015}. 

In the present paper we address to the 2D TIs since many experiments presently indicate the important role of structural defects in these materials, especially in the electron transport. Theoretical studies of bound states induced by nonmagnetic defects in the bulk of 2D TIs~ have revealed specific properties of these states which are inherent to 2D TIs and absent in the topologically trivial crystals.

It turns out that in 2D TIs there are two mechanisms of the bound state formation in contrast to the trivial case where a bound state arises only as a quasiparticle (electron or hole) is localized in a quantum well produced by the defect. In 2D TIs, bound states are formed by the repulsive potential as well. These states are similar to the helical edge states at the boundary of the 2D TI. Their distinguishing feature is that the electron density is low in the center but concentrated around the defect. Correspondingly there are two kinds of the states located at the defect. In particular, a defect with strongly localized potential induces two states irrespective of the sign of its potential. This contrasts to the trivial case where the same defect produces only one state. It is worth noting that two states located at a given defect are distinguished also by their pseudospin structure. In one state, the hole component of the spinor equals zero in the center while in the other state, the electron component turns to zero there. Correspondingly the states can be classified also as the electron-like and hole-like states~\cite{SablikovPSSR2014,SablikovPRB2015}.

In experiments, the defect-induced bound states are still poorly understood though they attract continuously increasing interest, which is stimulated mainly by discrepancies between the experiments and expectations of the theory. The discrepancies are usually associated with the presence of uncontrolled defects, the density of which is apparently high under the real conditions. 

In particular, the electric conductance of the edge states in experiments~\cite{KonigSciene2007,RothSciene2009,GusevKvonPRB2011,GrabeckiPRB2013,KnezPRL2014,LiPRL2015} appears to be smaller than the universal quantity $e^2/h$ predicted by the existing theories~\cite{HasanKaneRMP2010}. It turns out that in many experiments~\cite{GusevKvonPRB2011,GrabeckiPRB2013,KnezPRL2014,LiPRL2015} the conductance is decreased by several times or even orders of magnitude. Along with this, the common observation is that the suppression of the conductance weakly depends on the temperature. These facts suggests that the conductance reduction is related to structural defects in the TIs. Moreover, the energy spectrum of the defect-induced states plays an important role, since the scattering of electrons in helical edge states on non-magnetic defects is possible only due to inelastic processes~\cite{WuPRL2006,SchmidtPRL2012}, particularly due to intracenter transitions. Nevertheless, present theories of spin-flip scattering on defects are based on phenomenological models of a defect ignoring its real spectrum and electronic structure~\cite{VayrynenPRB2014,EssertPRB2015}. 

Our study is motivated by the idea that optical methods could be a promising tool for studying the unusual properties of defects in TIs. Such techniques are now well developed for defects in semiconductors~\cite{pajot2012optical} and are increasingly used in the studies of 2D electron systems~\cite{bonaccorso2010graphene,fang2013quantum}.

The present paper aims to clarify how the defect-induced states are manifested in optical absorption spectra. We find that the defects located in the bulk of the 2D TI produce specific peaks in the absorption spectrum which can be identified experimentally. The peaks originate from electron transitions between the quantum states of different types that are located at the same defect. In one state, a particle is captured by the attractive potential and in the other state, a particle circulates around the defect. An important factor affecting the formation of the peaks is also a specific dependence of the energy of the bound states of the defect potential. The energy of both states turns out to be weakly dependent on the amplitude of the defect potential in a wide range of the potentials. It is for this reason the light quanta absorbed by the defects with different potentials have close energies thereby forming the peak.

\section{Model and general equations}
\label{Model}
 
Our study of optical transitions in 2D TIs with structural defects is based on the Hamiltonian proposed by Bernevig, Hughes and Zhang (BHZ)~\cite{BHZScience2006}. This model adequately describes single-particle states in the low-energy region with using four-component basis $(|E\!\uparrow\rangle,|H\!\uparrow\rangle,|E\!\downarrow\rangle,|H\!\downarrow\rangle)^T$, where $|E\!\uparrow\rangle$ and $|E\!\downarrow\rangle$ are superposition of electron and light-hole states with the moment projection $m_J=\pm 1/2$, and $|H\!\uparrow\rangle$ and $|H\!\downarrow\rangle$ are the heavy-hole states with $m_J=\pm 3/2$. If the spin-orbit interaction (SOI) due to structural inversion asymmetry and bulk inversion asymmetry is absent, the BHZ Hamiltonian is block diagonal:
\begin{equation}
 \hat{H}_0=
\begin{pmatrix}
 h(\hat{\mathbf{k}}) & 0\\
 0 & h^*(-\hat{\mathbf{k}})
\end{pmatrix}\,,
\label{Hamiltonian}
\end{equation} 
where $h(\mathbf{k})=(C-Dk^2)\mathbb{I}_{2\times 2}+d_{\mu}(\mathbf{k})\sigma^{\mu}$ with $d_{\mu}(\mathbf{k})=(Ak_x,-Ak_y,M-Bk^2)$; $\hat{\mathbf k}$ is the momentum operator and $\sigma^{\mu}$ are the Pauli matrices. $A, B, C, D, M$ are model parameters depending on the quantum-well width~\cite{BHZScience2006}. Of critical importance is the sign of the parameter $M B$ that defines the ordinary and inverted-band situations: $M B>0$ corresponds to the topological phase, $M B<0$ is topologically trivial case. 

The defects are represented by a localized potential $V(|\mathbf{r}-\mathbf{r}_j|)$. We suppose that the density of the defects is not very high so that they do not interact one with other and can be considered independently. Particularly, we do not consider defect clusters forming electron puddles, which are sometimes discussed in connection with the scattering of electrons in edge states. In the present paper we study optical absorption due to defects with short-range potential. This is a reasonable simplification which is justified since the dielectric constant of TIs is usually high~\cite{footnote}
and therefore one can expect that the long-range part of the defect potential is small.

Consideration of the defects with short-range potential is of considerable interest also for the reason that in this case the structure of the bound state spectrum weakly depends on the details of the potential shape and therefore one can expect the appearance of universal results.

In what follows, it is convenient to use dimensionless notations for the energy $\varepsilon=E/|M|$, the distance $\tilde{r}=r\sqrt{M/B}$, the wave vector $\tilde{k}=k\sqrt{B/M}$, and the potential $v=V/|M|$. For brevity we will drop the tilde. In addition, we introduce important dimensionless parameters 
\begin{equation}
 a=\frac{A}{\sqrt{MB}},\quad d_b=\frac{D}{B}\,,
\end{equation} 
which determine the band structure of electrons in homogeneous crystal in the BHZ model. Parameter $d_b$ determines the asymmetry of the electron and hole bands, and the parameter $a$ to a large extent determines the shape of the dispersion relation in the conduction and valence bands.

In order to study the optical absorption one should calculate the matrix elements of the electron transitions between different states localized on an impurity and between the impurity states and band states.

First consider the band states in absence of the SOI. Since the Hamiltonian (\ref{Hamiltonian}) has block-diagonal structure, the spin-up and spin-down sectors can be considered independently. The band states are characterized by the band index $\lambda=\pm$, which corresponds to the $c$- and $v$-bands, and the spin $s=\uparrow,\downarrow$.

The electron energy is independent of the spin. The dispersion relation in the $c$- and $v$-bands has the following form:
\begin{equation}
 \varepsilon_{\lambda,s,k}=d_bk^2+\lambda\sqrt{(1-k^2)^2+a^2k^2}\,.
\end{equation}

The band states in the spin-up sector are presented by the following spinor:
\begin{equation}
 \Psi_{\lambda,\uparrow,k}=C_{\lambda,\uparrow,k}
 \begin{pmatrix}
 1\\
 \dfrac{ak_{-}}{\lambda\sqrt{(1-k^2)^2+a^2k^2}-1+k^2}\\
\end{pmatrix}
e^{i\mathbf{k}\mathbf{r}}\,,
\end{equation}
where $k_{\pm}=k_x\pm i k_y$, $C_{\lambda,\uparrow,k}$ is normalization constant.

The band spectrum has the qualitatively different shape in the following three regions of the parameters $a$ and $d_b$. At $a>\sqrt{2(1+|d_b|)}$, the energy gap equals 2$|M|$ and the dispersion curves has the minimum in the $c$-band and the maximum in the $v$-band at $k=0$. At $a<\sqrt{2(1-|d_b|)}$, the dispersion curve has a mexican hat shape in both bands, with the gap being smaller than 2$|M|$. When $\sqrt{2(1-|d_b|)}<a<\sqrt{2(1+|d_b|)}$, only one of the bands has a mexican hat shape. 

We now turn to the states bound at the defect. They are found from the Schr\"odinger equation: $[\hat{H}_0+V(r)-E]\Psi=0$. For the spin-up block we have
\begin{equation}
 [\varepsilon-h(\hat{\mathbf{k}})]\Psi(\mathbf{r})=v(r)\Psi(\mathbf{r})\,,
\end{equation} 
where $\Psi(\mathbf{r})=(\psi_e(\mathbf{r}),\psi_h(\mathbf{r}))^T$.

This equation is easily solved in the case where the potential is localized in a region which is small compared with the characteristic
length scale of the wave function~\cite{SablikovPSSR2014}. Using the Fourier transform we arrive at a system of equations for spinor components at the defect, $\bar{\Psi}(r\approx 0)\equiv(\bar{\psi}_e,\bar{\psi}_h)^T$,
\begin{align}
\label{BSE_e}
 \left[1\!-\!\int\limits_{0}^{\infty}dk k v_k\frac{\varepsilon\!-\!1\!+\!(1\!-\!d_b)k^2}{\Delta(\varepsilon,k)}\right]\bar{\psi}_e&=0\\
 \label{BSE_h}
 \left[1\!-\!\int\limits_{0}^{\infty}dk k v_k\frac{\varepsilon\!+\!1\!-\!(1\!+\!d_b)k^2}{\Delta(\varepsilon,k)}\right]\bar{\psi}_h&=0\,,
\end{align}
where $\Delta(\varepsilon,k)=(\varepsilon-d_bk^2)^2-(1-k^2)^2-a^2k^2$ and $v_k$ is the Fourier transform of the potential $v(r)$
\begin{equation}
 v_k=\int\limits_0^{\infty}drrv(r)J_0(kr)\,,
\end{equation} 
with $J_0(z)$ being the Bessel function.

The bound state energies are determined by Eqs~(\ref{BSE_e}) and (\ref{BSE_h}). It is obvious that they have solutions in two cases:\\
1. Electron-like states: $\bar{\psi}_e\ne 0$, $\bar{\psi}_h=0$. The energy of these states $\varepsilon=\varepsilon_e(v)$ is determined by the equation
\begin{equation}
 \int\limits_{0}^{\infty}dk k v_k\frac{\varepsilon\!-\!1\!+\!(1\!-\!d_b)k^2}{\Delta(\varepsilon,k)}=1\,,
\end{equation} 
2. Hole-like states: $\bar{\psi}_e=0$, $\bar{\psi}_h\ne 0$ with the energy $\varepsilon=\varepsilon_h(v)$ that is determined by the equation
\begin{equation}
 \int\limits_{0}^{\infty}dk k v_k\frac{\varepsilon\!+\!1\!-\!(1\!+\!d_b)k^2}{\Delta(\varepsilon,k)}=1\,.
\end{equation}

In each case there is only one solution~\cite{SablikovPSSR2014}. Thus, at a given potential of the defect there are two bound states: electron-like and hole-like ones. They are described by the following wave functions:
\begin{equation}\label{PsiEPsiH}
 \Psi_e(\mathbf{r})=C_e
 \begin{pmatrix}
 \psi_{1e}(r)\\
 i e^{-i\varphi}\psi_{2e}(r)
\end{pmatrix},
\quad
\Psi_h(\mathbf{r})=C_h
 \begin{pmatrix}
 i e^{i\varphi}\psi_{1h}(r)\\
 \psi_{2h}(r)
\end{pmatrix},
\end{equation} 
where $\varphi$ is the azimuth angle of $\mathbf{r}$,
\begin{align}
\label{psi1e}
 \psi_{1e}=&\int_0^{\infty}\!dkkv_k\frac{\varepsilon_e\!-\!1+\!(1\!-\!d_b)k^2}{\Delta(\varepsilon_e,k)}J_0(kr)\,,\\
 \label{psi2e}
 \psi_{2e}=&\int_0^{\infty}\!dkv_k\frac{ak^2}{\Delta(\varepsilon_e,k)}J_1(kr)\,,\\
 \label{psi1h}
 \psi_{1h}=&\int_0^{\infty}\!dkv_k\frac{ak^2}{\Delta(\varepsilon_h,k)}J_1(kr)\,,\\
 \label{psi2h}
 \psi_{2h}=&\int_0^{\infty}\!dkkv_k\frac{\varepsilon_h\!+\!1\!-\!(1\!+\!d_b)k^2}{\Delta(\varepsilon_h,k)}J_0(kr)\,.
\end{align}

\begin{figure}[ht]
\centerline{\includegraphics[width=.9\linewidth]{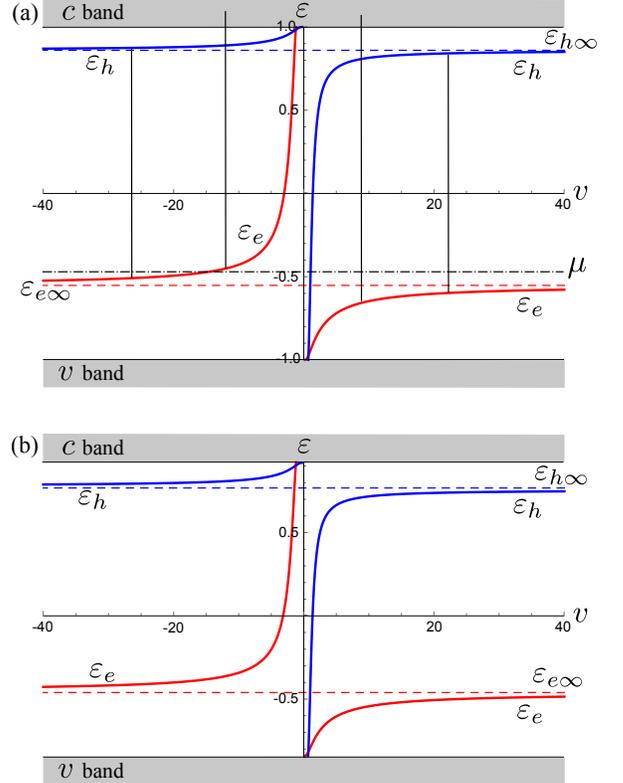}}
\caption{(Color online.) Energy of the electron-like, $\varepsilon_e$, and hole-like, $\varepsilon_h$, bound states as a function of the defect potential.      Dashed lines are the limiting energies $\varepsilon_{e\infty}$ and $\varepsilon_{h\infty}$, dash-and-dot line is the Fermi level $\mu$. (a) The calculation within the asymmetric BHZ model without the SOI. The arrows indicate the optical transitions, which are considered in the calculation of the absorption spectrum in Sec.~\ref{Transitions}. Calculations were carried out for $a$=2, $d_b$=0.2, $\Lambda$=5. (b) The same as in the panel (a), but with account of the SOI. The SOI parameter is $\Delta/|M|$=0.5.}
\label{fig1}
\end{figure}

To analyze possible spectra of the the defect-induced optical absorption, it is important to clarify how the energy of the bound states depends on the defect potential amplitude. In the case of 2D TIs, this dependence turns out to be very unusual. With increasing the potential, the energies of both the electron-like and hole-like states, $\varepsilon_e$ and $\varepsilon_h$, tend to finite limiting values, $\varepsilon_{e\infty}$ and $\varepsilon_{h\infty}$, as it is illustrated in Fig.~\ref{fig1}(a) for the potential of the form: $v(r)=v(\Lambda^2/\pi)\exp(-\Lambda^2r^2)$. The presence of these limiting energies of the bound states is a specific property of 2D TIs. 

The analysis of the bound states in a wide range of the parameters $a$ and $d_b$ shows that the properties of the bound states desribed above are qualitatively the same for all areas of the parameters. The only essential condition is the band structure inversion ($MB>0$).

The existence of the electron-like and hole-like bound states in 2D TIs as well as occurrence of the limiting energies was first found in our recent paper~\cite{SablikovPSSR2014} in the frame the BHZ model symmetric with respect to the electron and hole states. In this paper we generalize these results to the case of the asymmetric electron-hole band structure. It should be noted that the electron-hole asymmetry is very essential for all realistic 2D TIs used in experiments. 

Now we consider the effect of the SOI, which is also present in the realistic systems. To be specific we take into account the SOI arising due to both the bulk and interface inversion asymmetry, which is described by the Hamiltonian~\cite{KonigJPSJpn2008,DurnevPRB2016}
\begin{equation}
 H_{SOI}=
 \begin{pmatrix}
 0 & 0 & 0 & -\Delta\\
 0 & 0 & \Delta & 0\\
 0 & \Delta & 0 & 0\\
 -\Delta & 0 & 0 & 0
\end{pmatrix}.
\end{equation} 

We have studied this case within the same approach as above. In this case the calculations are much more cumbersome since the Hamiltonian does not split into two independent blocks and therefore the equations for eigenfunctions should be used in full 4$\times$4 form. However, the results turn out to be qualitatively very similar to those shown above. Therefore, we do not give the details, and present only the main results. 

We have found that a defect with short-range potential produces two bound states. Their energies vary with the potential amplitude in the same manner as in the case without the SOI. This is illustrated by Fig.~\ref{fig1}(b) where specific results are presented for the same parameters $a$ and $d_b$ as in Fig.~\ref{fig1}(a), but the SOI is added. It is seen that the SOI of rather high value ($\Delta/|M|$=0.5) does not qualitatively change the bound-state spectrum and its dependence on the defect potential. The main effects of the SOI are as follows: \\
\indent  i) The band gap decreases, but not much. In the case of Fig.~\ref{fig1}, the gap is decreased from 2 to $\approx$1.77.\\
\indent  ii) The spin state is changed. The $z$-component of the moment is not a quantum number. The moment deviates from the $z$-axis and depends on the potential amplitude.

Thus, the basic properties of the impurity states, which are important for our study, remain qualitatively unchanged. Therefore in the following calculations we drop the SOI.

The optical transitions between the bound states of different types and between the bound states and the band states are studied in the electric dipole approximation. Within the frame of the $\mathbf{kp}$-theory, the Hamiltonian of a light-matter interaction for spin-up sector reads
\begin{equation}\label{H_I}
 H_I=\frac{e}{\hbar c}\sqrt{\frac{B}{M}}
 \begin{pmatrix}
 2(1+d_b)\mathcal{A}\cdot\hat{\mathbf{k}} & a\mathcal{A}_+ \\
 a\mathcal{A}_- & -2(1-d_b)\mathcal{A}\cdot\hat{\mathbf{k}}
\end{pmatrix},
\end{equation} 
where 
\begin{equation}
 \mathcal{A}=\frac{e}{\hbar c}\sqrt{\frac{B}{M}}\,\mathbf{A}\,,
\end{equation} 
with $\mathbf{A}$ being the vector potential. Specific calculations are carried out for right-hand-polarized electromagnetic field incident normally on the sample.

\section{Optical transitions}
\label{Transitions}
In this section, we find out how the properties of the defect-induced states are manifested in the optical absorption spectra of 2D TIs. The main features of the absorption spectrum are caused by the intracenter transitions and the unusual dependence of the bound state energies on the defect potential. 

First of all, we note that the transitions between the electron-like and hole-like states are allowed in the electric dipole approximations in each spin sector without changing the spin since the orbital angular momenta of these states differ by unity. This conclusion follows from the direct calculation of the matrix elements $\langle h|H_I|e\rangle$ of the transition between $\Psi_e$ and $\Psi_h$ states. For example, the matrix element of the transition with the absorption of the light quantum on the defects with positive $v$ is expressed via the spinor components $\psi_{1e}$, $\psi_{2e}$, $\psi_{1h}$ and  $\psi_{2h}$ as follows 
\begin{multline}
\label{hAe}
\langle h|H_I|e\rangle\big|_{v>0}=C_h^*C_e\mathcal{A}_0\int\limits_0^{\infty}drr\biggl[a\psi_{2h}\psi_{1e}\\
\left.-(1\!+\!d_b)\psi_{1h}\frac{\partial \psi_{1e}}{\partial r}\!-\!(1\!-\!d_b)\psi_{2h}\left(\frac{\partial \psi_{2e}}{\partial r}\!+\!\frac{\psi_{2e}}{r}\right)\right]. 
\end{multline} 
A similar equation can also be obtained for defects with $v<0$. Using Eqs~(\ref{psi1e})-(\ref{psi2h}) it is easy to see that the right-hand side of Eq.~(\ref{hAe}) is nonzero and the matrix element $\langle h|H_I|e\rangle$ is a smooth function of $v$.

When calculating the probability of the optical transitions in the 2D TI, we assume that the real crystals of TIs contain a variety of point defects with very different potentials. The distribution of the defect potentials is described by a function $\rho(v)$, which is supposed to be a smooth function going to zero at $v\to 0$. We do not have actual data on the magnitude of the defect potentials in 2D TIs. Nevertheless, one can do some estimates taking into account that the potential is determined by the difference in the electron affinities of the defect and the host crystal. The electron affinity is usually of the order of Volts while the band gap in 2D TIs is typically of the order of 10~mV. Therefore, we can assume that the scatter in the defect potentials far exceeds the energy gap.

The rate of the transition with the absorption of photons is determined in the standard way~\cite{chuang2012physics} and has the form: 
\begin{multline}
 R(\omega)=\frac{4\pi}{\hbar}\int\limits_{-\infty}^{\infty}dv\rho(v)\bigl|\langle h|H_I|e\rangle\bigr|^2(f_e-f_h)\\
 \times\delta\left(\varepsilon_h(v)-\varepsilon_e(v)-\hbar\omega\right),
\end{multline}
where $\hbar\omega$ is the photon energy, $f_{e}$ and $f_{h}$ are the occupation probabilities of the electron-like and hole-like states. This equation can be simplified by the integration of the $\delta$-function:
\begin{equation}
\label{rate}
 R(\omega)=\frac{4\pi}{\hbar}\sum\limits_{j}\left[\rho(v)\bigl|\langle h|H_I|e\rangle\bigr|^2\frac{f_e\!-\!f_h}{\biggl|\frac{d(\varepsilon_h\!-\!\varepsilon_e)}{dv}\biggr|}\right]_{v=v_j},
\end{equation} 
where $v_j$ is a root of the equation
\begin{equation}
\label{v_j}
 \varepsilon_h(v)-\varepsilon_e(v)=\hbar\omega\,,
\end{equation} 
and the summation is over all roots $v_j$. 

\begin{figure}
\centerline{\includegraphics[width=.85\linewidth]{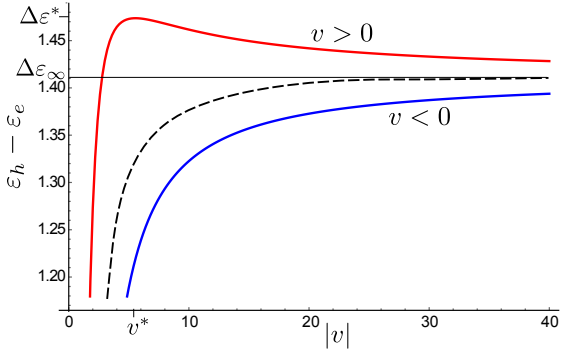}}
\caption{(Color online.) Energy difference of the hole-like and electron-like states, $\varepsilon_h-\varepsilon_e$, as a function of the defect potential.      Full lines correspond to positive and negative potential in the case where $d_b>0$. At $d_b<0$ the lines are swapped. Dashed line corresponds to the symmetric case, $d_b=0$. Calculations were carried out for $a$=2, $d_b$=0.2, $\Lambda$=5.}
\label{fig2}
\end{figure}

In general case, Eq.~(\ref{v_j}) has two roots. This is illustrated in Fig.~\ref{fig2} where the left-hand side of Eq.~(\ref{v_j}) is presented as a function of $v$. There are two branches of $[\varepsilon_h(v)-\varepsilon_e(v)]$ corresponding to the positive and negative potential. One of the branches (the branch for which $vd_b>0$) has a maximum $\Delta\varepsilon^*$ at $v=v^*$. In the limit $v\to \infty$, both branches go to the unique value $\Delta\varepsilon_{\infty}$. When $\Delta\varepsilon_{\infty}<\hbar\omega<\Delta\varepsilon^*$, there are two roots on the branch that has the maximum. At $\hbar\omega<\Delta\varepsilon_{\infty}$, there are also two roots, but they belong to different branches. Of course, in the other spin sector, the situation is the same. Eq.~(\ref{rate}) takes into account both spin sectors.

\begin{figure}[ht]
\centerline{\includegraphics[width=.9\linewidth]{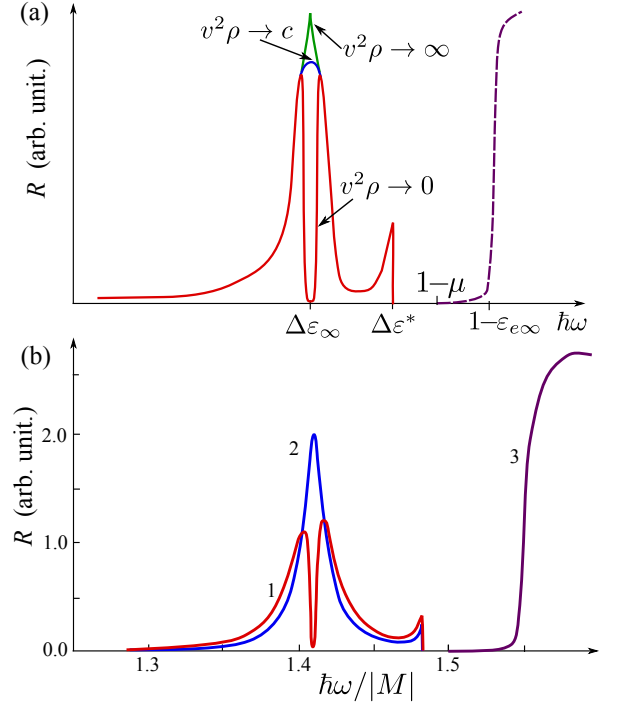}}
\caption{(Color online.) The spectrum of optical transition rate in 2D TIs with point defects. (a) Illustration to the qualitative analytical calculations (see the text). Lines 1, 2 and 3 describe the spectrum shape near the singularity point $\hbar\omega=\Delta\varepsilon_{\infty}$ for various asymptotics of the distribution function $\rho(v)$ at $v\to\infty$. Dashed line -- the absorption due to the electron transitions from the bound states into the conduction band. (b) The spectra calculated numerically for distribution function of different forms: 1 --  Gaussian, 2 -- Lorentzian, with $v_c$=50. Line 3 is the edge of the absorption due to the electron transitions from the defects to the conduction band. In this curve, $R$ is scaled 1:100. The parameters used in the calculation are $a$=2.0, $d_b$=0.2, $\Lambda$=5, $v_c$=50.}
\label{fig3}
\end{figure}

Equation~(\ref{rate}) allows one to qualitatively analyze main features of the absorption spectrum caused by the intracenter electron transitions. It is evident that a singularity of $R(\omega)$ appears when $\frac{d(\varepsilon_h\!-\!\varepsilon_e)}{dv}=0$. As can be seen from Fig.~\ref{fig2}, the singularity occurs at two points: $v_j=v^*$ and $v_j\to\infty$. Correspondingly there are two features in the spectrum of $R(\omega)$. At $\hbar\omega=\Delta\varepsilon^*$, the transition rate $R(\omega)$ has the singularity of the form $R\sim (\Delta\varepsilon^*-\hbar\omega)^{-1/2}$. Another singularity at $\hbar\omega=\Delta\varepsilon_{\infty}$ can be much more strong since the denominator in Eq.~(\ref{rate}) goes to zero as $(\hbar\omega-\Delta\varepsilon_{\infty})^2$. However, when $v\to \infty$, the distribution function goes to zero. Therefore the behavior of $R(\omega)$ near the singularity point is determined by the asymptotics of the ratio $\rho(v)/(\hbar\omega-\Delta\varepsilon_{\infty})^2$ at $v\to \infty$. It is easy to show that the denominator is proportional to $v^2$ as $v\to \infty$. Thus, $R(\omega)$ is finite in the singularity point if $\rho(v)\simeq v^{-2}$, $R(\omega)$ diverges if $\rho(v)$ decreases with $v$ slower than $v^{-2}$, and  $R(\omega)$ goes to zero if $\rho(v)$ decreases faster than $v^{-2}$.
This qualitative form of the absorption spectrum is illustrated in Fig.~\ref{fig3}(a), where we show the possible shapes of the spectrum near the singularity points ($\hbar\omega=\Delta\varepsilon^*$ and $\hbar\omega=\Delta\varepsilon_{\infty}$) for different distribution functions.

The full form of the spectrum of $R(\omega)$ is studied with the use of numerical calculations taking into account the electron transitions depicted by arrows in Fig.~\ref{fig1}(a), including the transitions from the defects to the $c$-band. The distribution function is taken in two forms: the Gaussian $\rho(v)=(v_c\sqrt{2\pi})^{-1}\exp[-v^2/(2v_c^2)]$ and the Lorentzian $\rho(v)=\pi^{-1}v_c/(v^2+v_c^2)$. The results are presented in Fig.~\ref{fig3}(b). They confirm generally the above qualitative picture. In the case of the Gaussian $\rho(v)$, the singularity at $\hbar\omega=\Delta\varepsilon_{\infty}$ has the form of two peaks divided by a deep crevasse, while in the case of the Loretzian $\rho(v)$ only one peak exists at this point. The shape of the spectrum near the other singularity at $\hbar\omega=\Delta\varepsilon^*$ has a form of an asymmetric peak in both cases. 

\section{Conclusion}
\label{Conclusion}

Electronic states bound at the structural defects in bulk of the 2D TIs are substantially different from the impurity states in trivial crystals. We have shown that the nontrivial properties of these states are manifested in the optical absorption spectrum. The characteristic features of the spectrum originate from the intracenter electron transitions.

We have studied in detail the case where the defects produce a short-range potential. These defects create two bound states with different pseudospin structure. The specific feature of 2D TIs is that the energy of the bound states depends on the defect potential in highly unusual way. The energies of both states tend to the limiting values with increasing the potential. It is important that electric dipole transition can occur between the states localized at the same defect and hence the intracenter electron transitions can largely determine the absorption of light. We have calculated the optical absorption spectrum taking into account another important fact that the real crystal contains a large number of different defects. We have described the variety of defects by the distribution function over the potential amplitude. 

The main result of our studies is that the defects lead to the formation of two singularities in the absorption spectrum. One singularity occurs at the energy close to the difference between the limiting values of the energies of the hole-like and electron-like bound states. The form of the spectrum near this point depends on the distribution function of the defect potentials. The singularity can has the form of a peak or two peaks divided by a deep crevasse. The second singularity is a peak. Its amplitude and position are determined by the asymmetry of the electron and hole bands in 2D TI.

Thus, we conclude that the optical spectroscopy of the bound states localized at defects can provide valuable information about nontrivial properties of TIs.

\section*{Acknowledgments}
This work was partially supported by Russian Foundation for Basic Research (Grant No~14-02-00237) and Russian Academy of Sciences.



\end{document}